\documentclass[aip, amsmath, amssymb,reprint,longbibliography,raggedbottom]{revtex4-2}

\usepackage[utf8]{inputenc}
\usepackage{graphicx}
\usepackage{dcolumn}
\usepackage{bm}
\usepackage[utf8]{inputenc}
\usepackage[T1]{fontenc}
\usepackage{mathptmx}
\usepackage{gensymb}
\usepackage{upgreek}
\usepackage{xcolor}
\usepackage{textcomp}
\usepackage{hyperref}
\usepackage{mhchem}
\usepackage{float}
\usepackage[verbose]{placeins}
\usepackage{todonotes}
\hypersetup{colorlinks=true,linkcolor=red,citecolor=blue,filecolor=magenta,urlcolor=magenta}
\usepackage{xr}
\usepackage{color,soul}

\usepackage{titlesec}
\titleformat{\section}[hang]{\small\bfseries\sffamily}{\thesection.}{0.5em}{\MakeUppercase}
\titlespacing{\section}{0pc}{1pc}{0.2pc}

\makeatletter
\def\frontmatter@thefootnote{*}
\def\frontmatter@makefnmark{%
  \@textsuperscript{\normalfont\@thefnmark}}
\makeatother

\begin{document}
\title{Rubidium-Referenced Frequency-Doubled Pockels Laser with High-Speed Electro-Optic Tuning}

\author{Xu Chen}%
\affiliation{School of Instrumentation and Optoelectronic Engineering, Beihang University, Beijing 100191, China}
\affiliation{Department of Materials Science and Engineering, National University of Singapore, 117575, Singapore}
\author{Michael Li}
\affiliation{Centre for Quantum Technologies, National University of Singapore, 117543, Singapore}
\author{Xiaodong Shi}
\affiliation{Centre for Quantum Technologies, National University of Singapore, 117543, Singapore}
\author{Lin Zhou}
\affiliation{Department of Materials Science and Engineering, National University of Singapore, 117575, Singapore}
\author{Ran Yang}
\affiliation{Centre for Quantum Technologies, National University of Singapore, 117543, Singapore}
\author{Mengyao Zhao}
\affiliation{Department of Materials Science and Engineering, National University of Singapore, 117575, Singapore}
\author{Veerendra Dhyani}
\affiliation{A$^\ast$STAR Quantum Innovation Centre (Q.InC), Agency for Science, Technology and Research (A$^\ast$STAR), 138634, Singapore}
\author{Sakthi Sanjeev Mohanraj}
\affiliation{Department of Materials Science and Engineering, National University of Singapore, 117575, Singapore}
\author{Guangxing Wu}
\affiliation{Department of Materials Science and Engineering, National University of Singapore, 117575, Singapore}
\author{Hao Hao}
\affiliation{Centre for Quantum Technologies, National University of Singapore, 117543, Singapore}
\author{Xia Yu}
\thanks{Email: Xia Yu (xiayu@buaa.edu.cn), Di Zhu (dizhu@nus.edu.sg)}
\affiliation{School of Instrumentation and Optoelectronic Engineering, Beihang University, Beijing 100191, China}
\affiliation{Hangzhou International Innovation Institute, Beihang University, Hangzhou 311115, China}
\author{Di Zhu}
\thanks{Email: Xia Yu (xiayu@buaa.edu.cn), Di Zhu (dizhu@nus.edu.sg)}
\affiliation{Department of Materials Science and Engineering, National University of Singapore, 117575, Singapore}
\affiliation{Centre for Quantum Technologies, National University of Singapore, 117543, Singapore}
\affiliation{A$^\ast$STAR Quantum Innovation Centre (Q.InC), Agency for Science, Technology and Research (A$^\ast$STAR), 138634, Singapore}

%%%%%%%%%%%%%%%%%%%%%%%%%%%%%%%%%%%%%%
\begin{abstract}

Narrow-linewidth lasers with rapid frequency control are essential for atomic physics and quantum technologies, yet combining these capabilities in chip-integrated sources at visible wavelengths remains challenging. Here we demonstrate a hybrid integrated InP/thin-film lithium niobate Pockels laser that emits at 780 nm through intracavity second-harmonic generation and supports gigahertz-bandwidth frequency tuning. A periodically poled lithium niobate waveguide is embedded in a laser cavity terminated by an extended distributed Bragg reflector, which provides electro-optically tunable feedback only at the 1560 nm fundamental wavelength. This design yields a singly resonant cavity that enhances frequency conversion while allowing the second-harmonic output to follow rapid tuning without resonance alignment. The laser achieves intrinsic linewidths of 2.7 kHz at 1560 nm and 9.0 kHz at 780 nm, with electro-optic tuning bandwidths beyond 1 GHz at both wavelengths. A mode-hop-free tuning range of 6.48 GHz at 780 nm enables direct spectroscopy of the rubidium D2 transition. Locking the second-harmonic output to this transition reduces the standard deviation of its frequency fluctuations from 135 MHz to 8 MHz and transfers the atomic frequency reference to the fundamental. These results establish integrated frequency-doubled Pockels lasers as compact, rapidly tunable sources that could simplify the optical architectures for future atomic and quantum photonic systems.
\end{abstract}
\maketitle

%%%%%%%% Body section %%%%%%%%%%%%

\section{Introduction}

Frequency-agile, narrow-linewidth lasers for addressing atomic transitions underpin applications ranging from neutral-atom quantum processors~\cite{evered_high-fidelity_2023,bluvstein_logical_2024} and atomic sensors~\cite{kitching_chip-scale_2018,stray_quantum_2022} to optical clocks~\cite{martin_compact_2018,hilton_mobile_2025} and precision spectroscopy~\cite{ferrari_precision_2003,morgenweg_ramsey-comb_2014}. 
Lasers around 780~nm are particularly important, as they access the rubidium (Rb) D2 transition used for cooling~\cite{sane_11_2012}, state preparation\cite{wangUltrafastHighFidelityState2025}, and frequency stabilization across cold-atom and vapor-cell technologies~\cite{{guan_780nm_2025,isichenko_multi-laser_2026}}. 
These applications demand not only narrow linewidth but also rapid frequency tuning, from gigahertz-scale frequency sweeps to high-bandwidth laser locking.  This combination is conventionally achieved with assemblies of benchtop semiconductor or frequency-doubled fiber lasers, external modulators, and bulk frequency-reference optics, whose size, power consumption, and system complexity remain bottlenecks to scalable deployment of quantum systems beyond the laboratory. Integrating these laser components on chip is therefore a key step toward compact atomic-photonic systems.

Integrated 780 nm lasers based on direct semiconductor gain have made impressive progress~\cite{luEmergingIntegratedLaser2024b}. GaAs/AlGaAs has been combined with low-loss photonic circuits through heterogeneous integration~\cite{zhangPhotonicIntegrationPlatform2023a,castro_integrated_2025,thiel2026wafer} or self-injection locking (SIL) to high-quality-factor (Q) resonators\cite{isichenko_sub-hz_2024,prokoshin_ultra-narrow-linewidth_2024}, producing chip-scale lasers with outstanding coherence, tunability, and stability. Frequency tuning in most of these lasers relies on thermo-optic effects, with typical bandwidths in the tens of kilohertz range~\cite{corato-zanarella_widely_2023,winkler_widely_2024}. 
Moving towards megahertz-scale tuning bandwidth, piezoelectric actuation is a viable solution~\cite{siddharth2025narrow}.
Further into the gigahertz range, electro-optic (EO) tuning becomes necessary. Recently, a 780 nm self-injection-locked laser has been demonstrated using an EO-tunable AlN resonator, achieving a tuning bandwidth approaching 1 GHz with a mode-hop-free tuning range of 2 GHz~\cite{xie_aluminum_2025}.
However, EO-tunable 780 nm lasers remain largely underdeveloped compared with their telecom counterparts. At short wavelengths, increased Rayleigh scattering makes low-loss waveguides and high-Q cavities more difficult to realize, with further complications such as increased photorefractive effects in EO material platforms. Semiconductor gain platforms are also less accessible than those at telecom wavelengths.

Nonlinear frequency conversion provides an alternative route, allowing 780 nm emission from mature telecom lasers through second-harmonic generation (SHG)~\cite{luntadila_tunable_2023}.
Hybrid-integrated 780 nm lasers have been demonstrated by self-injection locking a 1560 nm distributed feedback (DFB) semiconductor laser to a high-Q Si$_3$N$_4$ resonator, where the photogalvanic-induced all-optical poling allows SHG~\cite{li_high-coherence_2023,clementi_chip-scale_2023}. 
Thin-film lithium niobate (TFLN) offers a more natural platform for realizing integrated frequency-doubled lasers, combining strong intrinsic $\chi^{(2)}$ nonlinearity with flexible quasi-phase matching, high-speed EO modulation, and a wide transparency window~\cite{zhu_integrated_2021,lu_periodically_2019,snigirev_ultrafast_2023,shi2026integrated}. 
A high-Q periodically poled lithium niobate (PPLN) microresonator has been used to combine self-injection locking of a telecom DFB laser with resonance-enhanced SHG~\cite{ling_self-injection_2023}.
Using a different approach, an InP reflective semiconductor optical amplifier (RSOA) has been butt-coupled to a TFLN Vernier cavity formed by two racetrack resonators, one of which incorporates periodic poling for SHG and electrodes for fast EO tuning~\cite{li_integrated_2022}.
In the latter scheme, the fundamental wave (FW) can be tuned rapidly through EO modulation, but its second harmonic (SH) cannot follow synchronously, as efficient SHG requires precise resonance alignment to satisfy the double-resonance condition. Rapid frequency tuning at 780 nm therefore remains an outstanding challenge in integrated frequency-doubled lasers.

\begin{figure*}[htbp]
\centering
\includegraphics[width=0.99\textwidth]{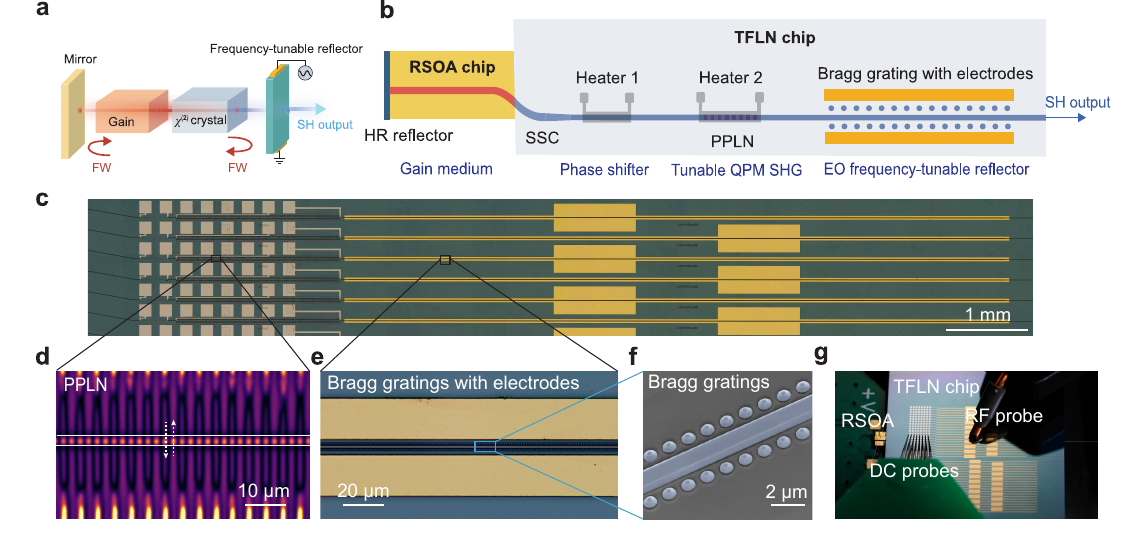}
\caption{\textbf{Design and implementation of the hybrid InP/TFLN frequency-doubled Pockels laser.}
\textbf{a,} Operating principle of the frequency-doubled Pockels laser. The fundamental-wave (FW) field is generated in the gain section and resonantly circulates between a mirror and an electrically tunable reflector, while the $\chi^{(2)}$ nonlinear crystal converts the intracavity FW field to the second-harmonic (SH) field.
\textbf{b,} Schematic of the hybrid InP/TFLN laser, consisting of an InP RSOA chip with a 
high-reflectivity (HR) back facet, a spot-size converter (SSC), a periodically poled lithium niobate (PPLN) waveguide with integrated heaters, and electro-optically tunable Bragg gratings with electrodes.
\textbf{c,} Optical micrograph of the fabricated TFLN photonic chip.
\textbf{d,} SH microscopic image of the PPLN waveguide. The bright sections show strong SHG response, while the dark lines in between mark the reconstructed domain walls.
\textbf{e,} Optical micrograph of the Bragg-grating section with electrodes.
\textbf{f,} Zoomed-in scanning electron micrograph of the Bragg gratings.
\textbf{g,} Photograph of the measurement setup, showing the RSOA chip, the direct-current (DC) probes applying power to the heater, and the radio-frequency (RF) probe applying an electric field to the electrodes.
}
\label{Fig1}
\end{figure*}

Here, we demonstrate a hybrid integrated InP/TFLN frequency-doubled Pockels laser that achieves gigahertz-bandwidth frequency tuning at both 1560 and 780 nm. Our design embeds a PPLN waveguide inside the laser cavity, terminated at one end by an EO-tunable extended distributed Bragg reflector (E-DBR). The E-DBR provides narrowband feedback at 1560~nm while remaining transmissive at 780~nm, making the cavity singly resonant at the FW. 
This scheme enables cavity-enhanced SHG while allowing rapid EO tuning without FW--SH resonance alignment.
The laser emits at 1560 and 780~nm with intrinsic linewidths of 2.7 and 9.0~kHz, respectively. At 780 nm, it delivers an on-chip power of 0.8~mW, tunes over a mode-hop-free range of 6.48~GHz, and responds to EO actuation with a 3-dB bandwidth of 1.2 GHz (Table~\ref{Table1}).
We stabilize the laser by locking the SH output to the Rb D2 transition, which transfers the atomic frequency reference to the FW at the same time. Our device may find applications in portable atomic clocks and quantum sensors, and the architecture can be extended to other atomic transitions by changing the PPLN and E-DBR designs.

\section{Results}

\subsection{Frequency-Doubled E-DBR Pockels Laser}

The frequency-doubled Pockels laser is based on an intracavity SHG architecture for simultaneous generation, tuning, and extraction of coherent FW and SH outputs (Fig.~\ref{Fig1}{a}).
It consists of a gain section, a broadband mirror, an electro-optically tunable frequency-selective reflector, and a $\chi^{(2)}$ nonlinear medium embedded inside the cavity.
The FW field oscillates within the cavity and builds up to a high intracavity power, driving SHG in the $\chi^{(2)}$ medium more efficiently than conventional single-pass frequency doubling~\cite{smith_theory_1970,nurmikko_intracavity_2003,flanniganStudyFundamentalWave2021}.
The tunable reflector is designed to provide feedback at the FW wavelength, while allowing the SH field to exit freely. 
Electrical actuation of the reflector tunes the cavity resonance and therefore the lasing frequency. 
Importantly, this singly resonant configuration avoids the stringent requirement of simultaneous FW and SH resonance alignment~\cite{li_integrated_2022, ling_self-injection_2023}, so that the SH output can follow the rapidly tuned FW .

The physical implementation of the laser is illustrated in Fig.~\ref{Fig1}{b}. 
An InP RSOA is butt-coupled to the TFLN chip through a spot-size converter (SSC) that reduces the mode mismatch between the RSOA and TFLN waveguide.
The TFLN chip integrates a PPLN waveguide for intracavity SHG, heaters for thermo-optic phase control and phase-matching tuning, and an EO-tunable E-DBR for wavelength-selective laser feedback and frequency tuning, as demonstrated in recent tunable E-DBR laser architectures~\cite{siddharth_piezoelectrically_2024,xue_pockels_2025,siddharth_ultrafast_2025,yu_resonator-enhanced_2026,larocque_sub-khz_2026}.

Figure~\ref{Fig1}{c} shows an optical micrograph of the device. 
The waveguide was fabricated on a 600-nm-thick x-cut TFLN film, with an etch depth of 300 nm. 
The PPLN waveguide has a width of 1 $\upmu$m, a length of 2 mm, and a poling period of 4.10 $\upmu$m for quasi-phase-matched SHG at 1560 nm. 
The SH microscopy image of the fabricated PPLN confirms uniform periodic poling (Fig.~\ref{Fig1}{d}). 
Two microheaters are integrated on the chip. One, above the PPLN section, provides thermo-optic tuning of the phase-matching condition, and the other, on the unpoled waveguide section, provides stable intracavity phase control. 

The E-DBR is designed as a weakly coupled grating for narrowband reflection at the FW wavelength, while remaining transmissive at its SH wavelength. 
The grating is 8 mm long with a period of 1.231 $\upmu$m (third-order grating for 1560 nm) and consists of 350-nm-diameter posts placed 750 nm away from the waveguide (Fig.~\ref{Fig1}f). 
The grating structure yields a weak coupling coefficient of $\kappa = 1.0 ~\mathrm{cm}^{-1}$ at 1560 nm, producing a narrow reflection band (see Supplementary Note 1 and Supplementary Figs.~1 and 2 for more details). 
The 780 nm SH lies outside the reflection band, and thus experiences negligible Bragg feedback when propagating through the grating section.
A pair of electrodes with a 9 $\upmu$m gap is placed around the E-DBR section, enabling high-speed EO modulation of the reflection band (Fig.~\ref{Fig1}e).

\subsection{Dual-Wavelength Lasing via Intracavity SHG}
Figure~\ref{Fig1}g shows the experimental setup for the laser operation. 
We butt-coupled an InP RSOA to the TFLN photonic integrated circuit and collected the output light using a lensed fiber. 
The 780 and 1560 nm outputs are separated by a fiber-based wavelength-division multiplexer (WDM) for independent characterization.

\begin{figure*}[!t]
\centering
\includegraphics[width=0.99\textwidth]{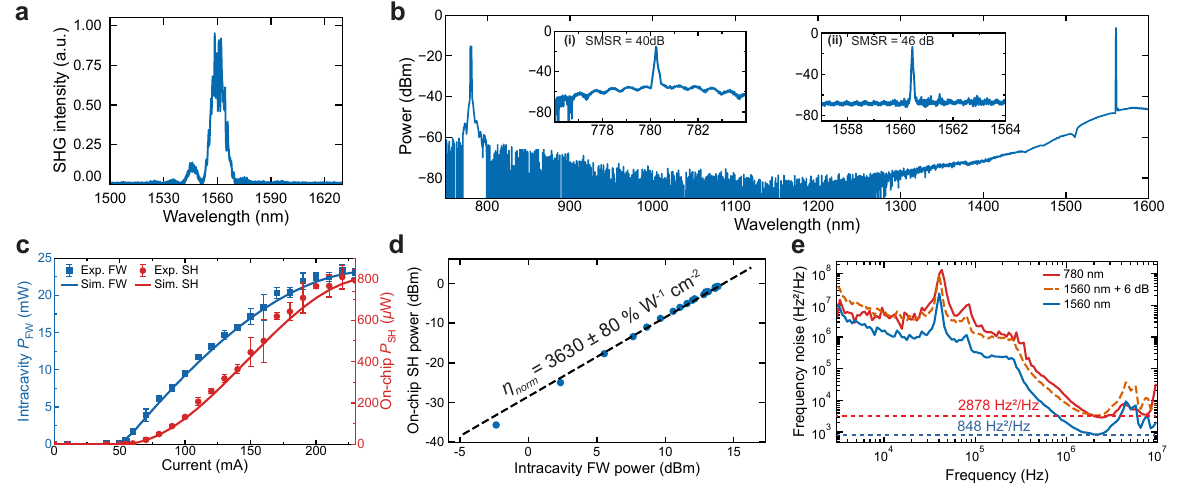}
    \caption{\textbf{Basic performance of the frequency-doubled Pockels laser.}
    \textbf{a,} Measured normalized SHG spectrum of the PPLN as a function of the FW wavelength, showing the wavelength-dependent conversion response.
    \textbf{b,} Measured optical spectrum of simultaneous lasing at 780 and 1560 nm. Insets show zoomed-in spectra around (i) 780 nm and (ii) 1560 nm, with side-mode suppression ratios of 40 dB and 46 dB, respectively.
     \textbf{c,} Measured and simulated intracavity FW power and on-chip SH power as functions of the RSOA injection current.
     \textbf{d,} On-chip SH power as a function of the intracavity FW power. The quadratic fit gives a normalized SHG efficiency of $3630~\pm~80\%~\mathrm{W}^{-1}\mathrm{cm}^{-2}$.
     \textbf{e,} Frequency-noise spectra of the 1560 and 780 nm outputs. The 1560 nm trace shifted up by 6 dB is also plotted for comparison.}
     \label{Fig2}
\end{figure*}

Before connecting the TFLN chip to the RSOA, we characterized the E-DBR reflection spectrum (see Supplementary Fig.~2) and the SHG spectrum of the PPLN using a tunable continuous-wave (CW) laser.
We aligned the phase-matching wavelength of the PPLN waveguide to the E-DBR reflection spectrum centered at 1560 nm (Fig.~\ref{Fig2}{a}) using the on-chip microheater. 
After butt-coupling the RSOA, we observed simultaneous laser emission at the visible and telecom wavelengths. 
The measured optical spectrum shows single-mode operation at both 780 and 1560 nm, with side-mode suppression ratios of 40 dB and 46 dB, respectively (Fig.~\ref{Fig2}{b}; captured using Yokogawa AQ6370E with a resolution of 0.02~nm). 

Figure~\ref{Fig2}{c} shows the measured and simulated intracavity FW power and on-chip SH power as functions of the RSOA injection current under CW operation. 
A lasing threshold current of around 55 mA was observed, and an on-chip optical power of 5.2 mW was achieved at 220\,mA, corresponding to an intracavity FW power of 23.4\,mW. 
At the same current, a stable on-chip SH power of 800\,$\upmu$W was obtained. 
The conversion efficiency, defined as the ratio of the on-chip SH power to the intracavity FW power, was estimated to be 3.4\%. 
Figure~\ref{Fig2}{d} plots the SH power as a function of the intracavity FW power, showing an SHG efficiency of $145~\pm~3$\%~W$^{-1}$ for the 2-mm-long PPLN waveguide, corresponding to a normalized SHG efficiency of $\eta_{\rm{norm}} = 3630~\pm~80$\%~W$^{-1}$~cm$^{-2}$.
To better understand the intracavity laser dynamics and the buildup of the SH field, we performed a traveling-wave simulation of the E-DBR laser with intracavity SHG (see Supplementary Note 2 and Supplementary Figs.~3--5 for details). 
The simulation reproduced the measured lasing threshold, the monotonic increase of the FW and SH powers, and the gradual saturation at high injection currents.

The spectral coherence of the 1560 and 780 nm outputs was quantified by frequency-noise measurements using a self-heterodyne method (Fig.~\ref{Fig2}{e}; see Methods)~\cite{yuan_correlated_2022}.
At high offset frequencies, the 1560 nm output reaches a white frequency-noise floor of 848 Hz$^2$/Hz, corresponding to an intrinsic linewidth of 2.7 kHz, and the 780 nm output reaches 2878 Hz$^2$/Hz, corresponding to 9.0 kHz. 
Since SHG doubles the optical phase fluctuations, the frequency-noise power spectral density of the SH is expected to be 6 dB higher than that of the FW.
The 1560 nm frequency-noise trace shifted upward by 6 dB is also plotted in Fig.~\ref{Fig2}{e}, showing reasonable agreement with the 780 nm trace.

\subsection{Electro-optic Frequency Tuning}

\begin{figure*}[!t]
\centering
\includegraphics[width=0.99\textwidth]{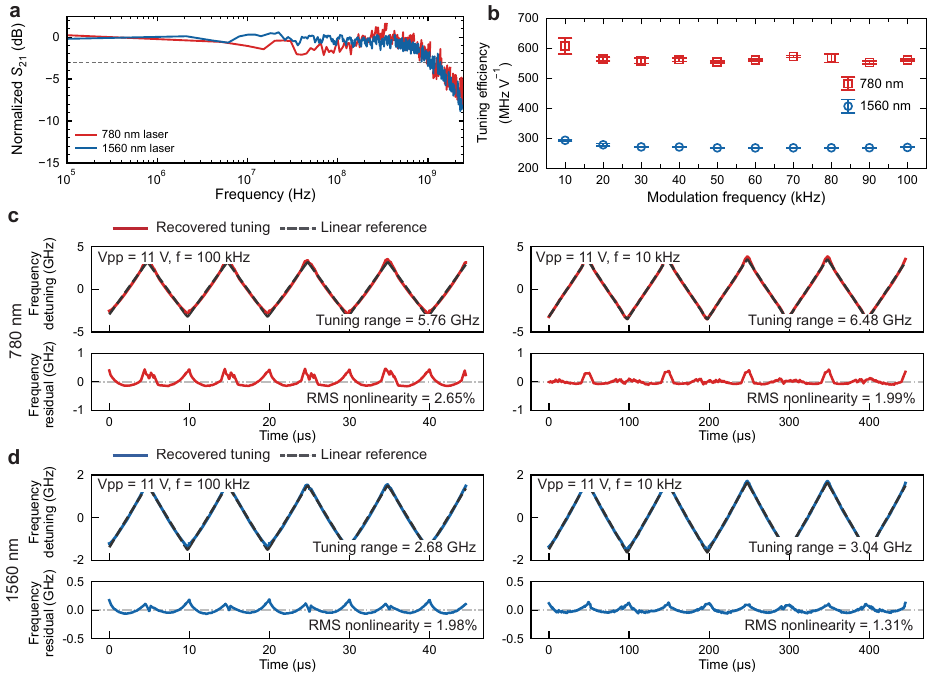}
\caption{\textbf{Electro-optic modulation of the frequency-doubled Pockels laser.}
\textbf{a,} Normalized $S_{21}$ tuning responses of the 780 and 1560 nm outputs. The dashed line marks the $-3$ dB level.
\textbf{b,} Tuning efficiency as a function of modulation frequency for the 1560 and 780 nm outputs. The approximately twofold ratio between the SH and FW tuning efficiencies is consistent with the nature of the frequency-doubling process.
\textbf{c,} Reconstructed frequency detuning and residual frequency error of the 780 nm output under triangular modulation with $V_{\mathrm{pp}} = 11$ V at 100 kHz (left) and 10 kHz (right).
\textbf{d,} Reconstructed frequency detuning and residual frequency error of the 1560 nm output under triangular modulation with $V_{\mathrm{pp}} = 11$ V at 100 kHz (left) and 10 kHz (right). 
Their mode-hop-free tuning ranges and root-mean-square (RMS) nonlinearities are given in the plots.}
\label{Fig3}
\end{figure*}

Electro-optic actuation of the E-DBR allows frequency tuning of both 1560 and 780 nm outputs with gigahertz bandwidth. 
The voltage applied across the E-DBR electrodes changes the refractive index of the TFLN waveguide and Bragg grating through the Pockels effect, shifting the cavity round-trip phase and the E-DBR reflection wavelength simultaneously. 
This actuation tunes the FW frequency, and the SH frequency follows synchronously.
We characterized the small-signal tuning bandwidth using a vector network analyzer (VNA) and an unbalanced Mach--Zehnder interferometer (uMZI) as a frequency discriminator. 
The measured $S_{21}$ tuning responses were corrected by removing the transfer function of the uMZI, and subsequently fitted with a first-order low-pass filter model to extract the bandwidths (see Supplementary Note 3 and Supplementary Figs.~6 and 7 for details). 
The extracted 3-dB tuning bandwidths are 1.11$~\pm~$0.03 GHz for the 1560 nm output and 1.2$~\pm~$0.1 GHz for the 780 nm output (Fig.~\ref{Fig3}a).

Under large-signal modulation, the laser achieved multi-gigahertz mode-hop-free tuning with a linear voltage-to-frequency response. 
We applied triangular voltage waveforms to the E-DBR electrodes and reconstructed the dynamic frequency detuning from the uMZI signal (see Supplementary Note 4 and Supplementary Fig.~8 for details).
At the largest applied peak-to-peak voltage that maintained mode-hop-free operation ($V_{\mathrm{pp}}=11~\mathrm{V}$), the recovered traces in Figs.~\ref{Fig3}{c} and {d} closely followed the corresponding ideal triangular reference waveforms.
The maximum measured mode-hop-free tuning ranges were 3.04 GHz for the 1560 nm output and 6.48 GHz for the 780 nm output, with corresponding RMS nonlinearities of 1.31\% and 1.99\%, respectively.

We further extracted the tuning efficiencies at different modulation frequencies and summarized them in Fig.~\ref{Fig3}{b}. 
Over the 20--100 kHz range, the tuning efficiencies remain at $268.8\pm0.4 $ MHz/V for the 1560 nm output and $561\pm4$ MHz/V for the 780 nm output. 
The SH tuning efficiency is approximately twice the FW tuning efficiency, consistent with the nature of the frequency-doubling process.
The small deviation from exact twofold scaling likely arises from the phase-to-frequency conversion in the independent uMZI-based reconstructions.
The slightly higher tuning efficiency observed at 10 kHz may be associated with relaxation-related charge redistribution in TFLN, an effect that has been reported to enhance the EO response over the 1--20 kHz range~\cite{holzgrafe_relaxation_2024}.
These results show that intracavity SHG preserves the FW tuning response, while extending rapid frequency control to the SH output.

\subsection{Frequency Referencing with Rb Vapor Cell}

\begin{figure*}[!t]
\centering
\includegraphics[width=0.99\textwidth]{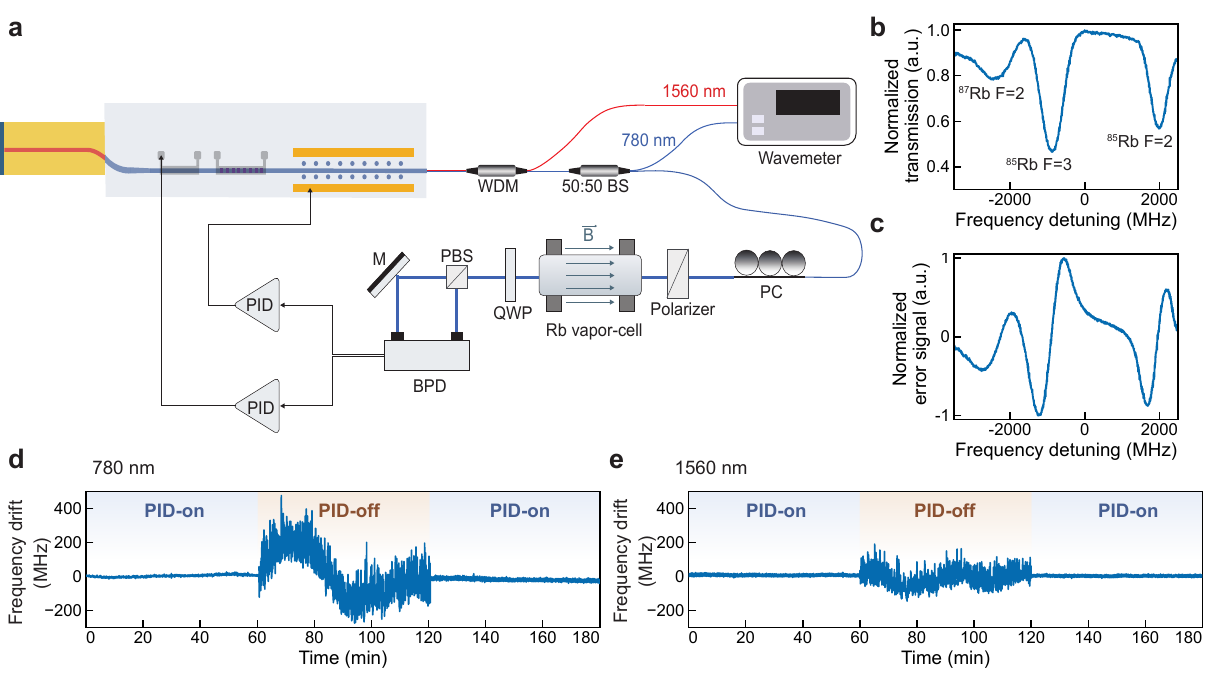}
\caption{\textbf{Laser frequency stabilization.}
\textbf{a,} Experimental setup for laser locking based on a dichroic atomic vapor laser lock (DAVLL). A WDM separates the 1560 nm and 780 nm outputs; both are monitored by a wavemeter, while the 780 nm SH output is sent through the Rb vapor-cell setup to generate feedback signals for the thermo-optic phase shifter and E-DBR electrodes. BS: beam splitter, PC: polarization controller, PBS: polarizing beam splitter, QWP: quarter-wave plate, M: mirror, BPD: balanced photodetector.
\textbf{b,} Measured Rb D2 transmission spectrum, with the main Doppler-broadened absorption features labeled by isotope and ground-state hyperfine level.
\textbf{c,} Normalized DAVLL error signal obtained from the differential absorption of the circular polarization components.
Recorded frequency drifts of \textbf{d,} 780 nm and \textbf{e,} 1560 nm outputs with the proportional--integral--derivative (PID) controllers enabled, disabled, and re-enabled.}
\label{Fig4}
\end{figure*}

Locking a laser to an atomic transition suppresses frequency drift and provides an absolute frequency reference for precision metrology and quantum applications. 
Here, we reference the 780 nm output to the Rb D2 transition using a dichroic atomic vapor laser lock (DAVLL)~\cite{corwin_frequency-stabilized_nodate}, thereby also stabilizing the 1560 nm output. 
The setup is shown in Fig.~\ref{Fig4}{a}, where the 1560 and 780 nm outputs were sent to a high-resolution wavemeter (HighFinesse WS7-60 IR-I, resolution of 2 MHz) for frequency monitoring. 
The 780 nm output was partially directed to the DAVLL setup, where it passed through a Rb vapor cell under a magnetic field. 
The magnetic-field-induced circular dichroism produces differential absorption of the two circular polarization components, which were analyzed using polarization optics and balanced detection to generate an error signal for laser locking. 
Figures~\ref{Fig4}b and c show the Rb D2 transmission spectrum, with the labeled main Doppler-broadened absorption features, and the DAVLL error signal. 
The lock point was the zero crossing associated with the $^{85}$Rb $F=3$ feature. 
The error signal was sent to two independent proportional--integral--derivative (PID) controllers to provide feedback signals to the thermo-optic phase shifter and the E-DBR electrodes for slow drift compensation and fast EO correction, respectively.

We also performed repeatable locking and relocking of the laser to the Rb D2 transition. 
The PID loop was first enabled for 60 min, then disabled for 60 min to record the free-running drift, and finally re-enabled for another 60 min to show the relocking process. 
The frequency fluctuations of the 780 and 1560 nm outputs are shown in Figs.~\ref{Fig4}{d} and {e}, respectively. 
Limited by available channels in the wavemeter, these two traces were recorded in separate measurements instead of concurrently.
For the 780 nm output, the free-running frequency fluctuation had a peak-to-peak range of 749 MHz and a standard deviation of 135 MHz. 
Under closed-loop stabilization, these values were reduced to 57 MHz peak-to-peak and a standard deviation of 8 MHz. 
A similar improvement was observed for the 1560 nm output, where the peak-to-peak fluctuation was reduced from 335 MHz to 32 MHz, and the standard deviation was reduced from 47 MHz to 4 MHz. 
The locked fractional-frequency Allan deviations at an averaging time of 1~s were $4.4\times10^{-9}$ and $5.1\times10^{-9}$ for the 1560 and 780~nm outputs, respectively (see Supplementary Note 5 and Supplementary Fig.~9 for more details).
The suppression of frequency fluctuations at both wavelengths confirms that locking the 780 nm output transfers the Rb atomic reference to the 1560 nm output through their coherent frequency relation.

\section{Discussion}

\begin{table*}[!t]
\centering

\makeatletter
\let\oldmakecaption\@makecaption
\long\def\@makecaption#1#2{%
  \par
  \vskip\abovecaptionskip
  \begingroup
    \normalsize\rmfamily
    \noindent\raggedright
    #1.\enspace#2\par
  \endgroup
  \vskip\belowcaptionskip
}
\makeatother

\caption{\textbf{Performance comparison with state-of-the-art integrated lasers around 780 nm}, including material platform, laser~architecture, tuning bandwidth, tuning range,
laser linewidth ($\Delta\nu$), tuning mechanism, output power
($P_{\mathrm{out}}$), and SHG efficiency
($\eta_{\mathrm{SHG}}$).}
\label{Table1}

\makeatletter
\let\@makecaption\oldmakecaption
\makeatother

\normalsize
\setlength{\tabcolsep}{1pt}
\renewcommand{\arraystretch}{1.15}

\def\tcell#1{%
  \begin{tabular}[c]{@{}c@{}}
  #1
  \end{tabular}%
}

\begin{tabular*}{\textwidth}{
@{\extracolsep{\fill}}
ccccccccc
@{}
}

\hline
\hline

\multicolumn{9}{c}{
\textbf{Integrated frequency-doubled lasers}
} \\

\hline
\noalign{\vskip 2pt}

\tcell{\textbf{Ref.}} &
\tcell{\textbf{Laser}\\\textbf{platform}} &
\tcell{\textbf{Laser}\\\textbf{architecture}} &
\tcell{\textbf{Tuning}\\\textbf{bandwidth}} &
\tcell{\textbf{Tuning}\\\textbf{range}} &
\tcell{\textbf{$\Delta\nu$}} &
\tcell{\textbf{Tuning}\\\textbf{mechanism}} &
\tcell{\textbf{$P_{\mathrm{out}}$}\\\textbf{(mW)}} &
\tcell{\textbf{$\eta_{\mathrm{SHG}}$}\\
($\%\,\mathrm{W}^{-1}$)}
\\

\noalign{\vskip 2pt}
\hline

\tcell{\textbf{This}\\\textbf{work}} &
\tcell{LiNbO$_3$\\(hybrid)} &
\tcell{External cavity\\(E-DBR)} &
\tcell{1.2 GHz} &
\tcell{6.48 GHz} &
\tcell{9.0 kHz} &
\tcell{EO} &
\tcell{0.8} &
\tcell{145}
\\

\tcell{[\onlinecite{li_high-coherence_2023}]} &
\tcell{Si$_3$N$_4$\\(hybrid)} &
\tcell{SIL (ring)} &
\tcell{--} &
\tcell{--} &
\tcell{25 Hz} &
\tcell{--} &
\tcell{$>0.5$} &
\tcell{$114$\textsuperscript{a}}
\\

\tcell{[\onlinecite{clementi_chip-scale_2023}]} &
\tcell{Si$_3$N$_4$\\(hybrid)} &
\tcell{SIL (ring)} &
\tcell{--} &
\tcell{--} &
\tcell{163 Hz\\(predicted)} &
\tcell{--} &
\tcell{2.3} &
\tcell{280}
\\

\tcell{[\onlinecite{ling_self-injection_2023}]} &
\tcell{LiNbO$_3$\\(hybrid)} &
\tcell{SIL (ring)} &
\tcell{--} &
\tcell{--} &
\tcell{4.7 kHz} &
\tcell{--} &
\tcell{$>2$} &
\tcell{3000\textsuperscript{a}}
\\

\tcell{[\onlinecite{li_integrated_2022}]} &
\tcell{LiNbO$_3$\\(hybrid)} &
\tcell{External cavity\\(Vernier)} &
\tcell{--} &
\tcell{$>10$ nm\\(coarse)} &
\tcell{--} &
\tcell{Thermo-optic\textsuperscript{b}} &
\tcell{--} &
\tcell{--}
\\

\hline

\multicolumn{9}{c}{
\textbf{Integrated lasers emitting directly}
} \\

\hline
\noalign{\vskip 3pt}

\tcell{\textbf{Ref.}} &
\tcell{\textbf{Laser}\\\textbf{platform}} &
\tcell{\textbf{Laser}\\\textbf{architecture}} &
\tcell{\textbf{Tuning}\\\textbf{bandwidth}} &
\tcell{\textbf{Tuning}\\\textbf{range}} &
\tcell{\textbf{$\Delta\nu$}} &
\tcell{\textbf{Tuning}\\\textbf{mechanism}} &
\multicolumn{2}{c}{
\tcell{\textbf{$P_{\mathrm{out}}$}\\\textbf{(mW)}}
}
\\

\noalign{\vskip 2pt}
\hline

\tcell{[\onlinecite{isichenko_sub-hz_2024}]} &
\tcell{Si$_3$N$_4$\\(hybrid)} &
\tcell{SIL (ring)} &
\tcell{--} &
\tcell{2 nm (coarse)\\2.5 GHz (fine)} &
\tcell{0.74 Hz} &
\tcell{Thermo-optic} &
\multicolumn{2}{c}{
\tcell{2\textsuperscript{c}}
}
\\

\tcell{[\onlinecite{prokoshin_ultra-narrow-linewidth_2024}]} &
\tcell{Si$_3$N$_4$\\(hybrid)} &
\tcell{SIL (ring)} &
\tcell{--} &
\tcell{--} &
\tcell{105 Hz\textsuperscript{d}} &
\tcell{--} &
\multicolumn{2}{c}{
\tcell{--}
}
\\

\tcell{[\onlinecite{corato-zanarella_widely_2023}]} &
\tcell{Si$_3$N$_4$\\(hybrid)} &
\tcell{SIL (ring)} &
\tcell{34 kHz} &
\tcell{12.5 nm (coarse)\\33.9 GHz (fine)} &
\tcell{$<1.1$ kHz} &
\tcell{Thermo-optic} &
\multicolumn{2}{c}{
\tcell{11.24}
}
\\

\tcell{[\onlinecite{xie_aluminum_2025}]} &
\tcell{AlN (hybrid)} &
\tcell{SIL (ring)} &
\tcell{1 GHz} &
\tcell{2 GHz} &
\tcell{200 kHz\textsuperscript{d}} &
\tcell{EO} &
\multicolumn{2}{c}{
\tcell{0.36}
}
\\

\tcell{[\onlinecite{corato-zanarella_lowswap_2026}]} &
\tcell{Si$_3$N$_4$\\(hybrid)} &
\tcell{SIL (ring)} &
\tcell{--} &
\tcell{9.5 nm (coarse)\\50 GHz (fine)} &
\tcell{0.7 kHz} &
\tcell{Thermo-optic} &
\multicolumn{2}{c}{
\tcell{30\textsuperscript{c}}
}
\\

\tcell{[\onlinecite{castro_integrated_2025}]} &
\tcell{Si$_3$N$_4$\\(heterogeneous)} &
\tcell{External cavity\\(Vernier)} &
\tcell{--} &
\tcell{17 nm (coarse)\\108 GHz (fine)} &
\tcell{5.85 kHz} &
\tcell{Thermo-optic} &
\multicolumn{2}{c}{
\tcell{$>10$}
}
\\

\hline
\hline

\end{tabular*}

\vspace{2pt}

\begin{minipage}{\textwidth}
\raggedright
\footnotesize

\textsuperscript{a} SHG efficiency measured using an external
continuous-wave laser.

\textsuperscript{b} The listed $>10$ nm tuning range corresponds to
coarse thermo-optic tuning of the visible output; EO actuation
was also demonstrated for frequency modulation of the fundamental wave
and switching of the visible output.

\textsuperscript{c} Off-chip output power. All other output powers are on-chip
values, as reported or estimated from the reported coupling loss.

\textsuperscript{d} Linewidth obtained from heterodyne or self-heterodyne
measurement while the other values denote intrinsic linewidths.

\end{minipage}

\end{table*}

We have demonstrated a hybrid InP/TFLN frequency-doubled Pockels laser that generates coherent 780 nm light through intracavity SHG in an E-DBR cavity. 
The device provides narrow-linewidth operation and fast EO frequency tuning at both the 1560 nm and 780 nm wavelengths.
Locking the 780 nm output to the Rb D2 transition suppresses long-term frequency drift and transfers the atomic reference to the coherently linked 1560 nm output.
A performance comparison with state-of-the-art integrated 780 nm lasers is provided in Table~\ref{Table1}, highlighting the combination of intracavity frequency doubling, narrow-linewidth operation, and gigahertz-scale EO frequency control achieved in the demonstrated device.

The present SH output power can be further increased by reducing intracavity and interface losses. 
The coupling loss at the RSOA--TFLN taper interface is $\sim$4 dB. 
Reducing this loss through better mode matching would increase the intracavity FW power, leading to higher absolute conversion efficiency.
For Rb-referenced operation, the residual frequency fluctuations may be partly attributed to fluctuations of the DAVLL error signal, which is sensitive to environmental perturbations such as magnetic-field variations. 
Incorporating environmental isolation and using a higher-sensitivity frequency discriminator, such as modulation-transfer spectroscopy~\cite{leeLaserFrequencyStabilization2023}, could enhance the frequency stability.
Further optimization in packaging, passive cavity stability, and EO feedback bandwidth could also improve the atom-referenced stabilization performance.

More broadly, the architecture is not restricted to 780 nm operation for Rb referencing, but offers a route to short wavelengths that are difficult to access with direct semiconductor gain, such as the green gap~\cite{moustakasOptoelectronicDevicePhysics2017}.
By combining different semiconductor gain bands with engineered quasi-phase-matching profiles, the same concept could be extended to other visible wavelengths addressing other atomic and molecular transitions.
In addition to its spectral versatility, the integration of mature telecom-band semiconductor gain with nonlinear frequency conversion and rapid EO frequency tuning could reduce reliance on separate wavelength-conversion and frequency-tuning modules in future atomic-photonic systems. This functional integration could simplify system design and support the development of more sustainable and scalable photonic technologies.

\section{Methods}
\subsection{Device fabrication}
The device was fabricated on a 600-nm-thick MgO-doped x-cut TFLN (NanoLN). The poling electrode was patterned by
electron-beam (e-beam) lithography, e-beam metal evaporation (120 nm Ni), and lift-off. 
Electrical poling was performed by applying a series of high-voltage pulses to reverse the ferroelectric domains in TFLN. 
The waveguide and Bragg grating structures were defined by e-beam lithography, followed by Ar$^+$ inductively coupled plasma reactive-ion etching (ICP-RIE) to a depth of 300 nm. 
We then deposited a 1.6 $\upmu \mathrm{m}$-thick SiO$_2$ cladding layer by plasma-enhanced chemical vapor deposition (PECVD). 
The microheaters were patterned using ultraviolet lithography, e-beam evaporation of 100 nm Pt, and lift-off. 
Next, the EO tuning electrodes were patterned by ultraviolet lithography, e-beam evaporation of 500 nm Cu and 50 nm Au, and lift-off. 
Finally, the chip facets were diced and polished for coupling with the gain chip.

\subsection{Power measurement and loss characterization}

For the current-dependent power measurements (Fig.~\ref{Fig2}c), the RSOA injection current was gradually increased, while the laser was operated in the CW regime. 
Since a small change in the driving current can shift the intracavity optical phase and induce mode hopping, the power applied to the thermo-optic phase shifter was adjusted during the current sweep to maintain single-mode operation. 

The intracavity FW power was calculated from the powers measured in the lensed fiber, E-DBR response, and chip-to-fiber coupling loss.
We measured the chip-to-fiber coupling loss at 1560~nm to be 5.5 dB using a neighboring reference waveguide fabricated on the same chip.
The propagation loss and the reflectivity of the E-DBR section were calibrated to be 5.0~dB and 0.3, respectively. 
Taking these values into account, the intracavity FW power is estimated through $P_{\mathrm{FW,cav}}=\frac{P_{\mathrm{FW,fib}}}{(1-R_{\mathrm{DBR}})10^{-(L_{\mathrm{DBR}}+L_{\mathrm{c,FW}})/10}}$,
where $P_{\mathrm{FW,fib}}$ is the measured output FW power in the lensed fiber, $P_{\mathrm{FW,cav}}$ is the estimated intracavity FW power, $L_{\mathrm{DBR}}$ denotes the estimated loss associated with the E-DBR section, and $L_{\mathrm{c,FW}}$ denotes the estimated chip-to-fiber coupling loss at 1560~nm.
The chip-to-fiber coupling loss at 780 nm was calibrated to be 5.0~dB from transmission measurements of the neighboring reference waveguide. 
This loss was used to infer the corresponding on-chip SH power from the measured output SH power in the lensed fiber.

\subsection{Laser frequency-noise measurement}
The frequency noise of the 1560 nm and 780 nm outputs was measured individually using the self-heterodyne method (Fig.~\ref{Fig2}e)~\cite{yuan_correlated_2022}. 
After the two laser wavelengths were separated by the fiber-based WDM, each output was directed to an MZI configured for the corresponding wavelength, with a 17-m fiber delay in one arm and an acousto-optic modulator (AOM) in the other. 
The AOMs used for the 780 nm and 1560 nm measurements were driven at 80 MHz and 100 MHz, respectively. 
The resulting heterodyne signal at the MZI output was detected by a photodetector and recorded with an oscilloscope. 
For each wavelength, the phase of the heterodyne carrier was extracted from the waveform using a Hilbert transform. 
The phase noise was further extracted and converted into the frequency noise, enabling the white-noise floor and corresponding intrinsic linewidth to be determined at each wavelength.

\begin{acknowledgments}
D.Z. acknowledges funding support by the Ministry of Education, Singapore, under its Academic Research Fund Tier 1 (FY2025). X.C. acknowledges funding support under the China Scholarship Council. X.Y. acknowledges funding support under the University International Collaboration Incubation Project (501GJHZ2026117002).

\end{acknowledgments}

\section*{Author Contributions}
D.Z., X.C., and X.S. conceived the idea.
X.C. designed the device.
X.C., X.S., R.Y., V.D., S.S.M., G.W., and H.H. fabricated the devices.
X.C., M.L., X.S., L.Z., R.Y., and M.Z. performed the measurements.
D.Z. and X.Y. supervised the project.
All authors discussed the results and wrote the manuscript.

\section*{Data Availability}
The data that supports the findings of this study is available from the
corresponding author on reasonable request.

\section*{Competing interests}
The authors declare no competing interests.

\section*{REFERENCES}

\bibliography{References_for_the_SHG_laser_checked_corrected}

@article{li_high-coherence_2023,
  author = {Li, Bohan and Yuan, Zhiquan and Jin, Warren and Wu, Lue and Guo, Joel and Ji, Qing-Xin and Feshali, Avi and Paniccia, Mario and Bowers, John E. and Vahala, Kerry J.},
  title = {High-coherence hybrid-integrated 780 nm source by self-injection-locked second-harmonic generation in a high-{Q} silicon-nitride resonator},
  journal = {Optica},
  volume = {10},
  number = {9},
  pages = {1241--1244},
  year = {2023},
  doi = {10.1364/OPTICA.498391},
  url = {https://opg.optica.org/abstract.cfm?URI=optica-10-9-1241},
}

@article{ling_self-injection_2023,
  author = {Ling, Jingwei and Staffa, Jeremy and Wang, Heming and Shen, Boqiang and Chang, Lin and Javid, Usman A. and Wu, Lue and Yuan, Zhiquan and Lopez-Rios, Raymond and Li, Mingxiao and He, Yang and Li, Bohan and Bowers, John E. and Vahala, Kerry J. and Lin, Qiang},
  title = {Self-Injection Locked Frequency Conversion Laser},
  journal = {Laser \& Photonics Reviews},
  volume = {17},
  number = {5},
  pages = {2200663},
  year = {2023},
  doi = {10.1002/lpor.202200663},
  url = {https://onlinelibrary.wiley.com/doi/abs/10.1002/lpor.202200663},
}

@article{prokoshin_ultra-narrow-linewidth_2024,
  author = {Prokoshin, Artem and Gehl, Michael and Madaras, Scott and Chow, Weng W. and Wan, Yating},
  title = {Ultra-narrow-linewidth hybrid-integrated self-injection locked laser at 780 nm},
  journal = {Optica},
  volume = {11},
  number = {7},
  pages = {1024--1029},
  year = {2024},
  doi = {10.1364/OPTICA.531152},
  url = {https://opg.optica.org/abstract.cfm?URI=optica-11-7-1024},
}

@article{siddharth_piezoelectrically_2024,
  author = {Siddharth, Anat and Attanasio, Alaina and Bianconi, Simone and Lihachev, Grigory and Zhang, Junyin and Qiu, Zheru and Bancora, Andrea and Kenning, Scott and Wang, Rui Ning and Voloshin, Andrey S. and Bhave, Sunil A. and Riemensberger, Johann and Kippenberg, Tobias J.},
  title = {Piezoelectrically tunable, narrow linewidth photonic integrated extended-{DBR} lasers},
  journal = {Optica},
  volume = {11},
  number = {8},
  pages = {1062--1069},
  year = {2024},
  doi = {10.1364/OPTICA.524703},
  url = {https://opg.optica.org/abstract.cfm?URI=optica-11-8-1062},
}

@article{li_integrated_2022,
  author = {Li, Mingxiao and Chang, Lin and Wu, Lue and Staffa, Jeremy and Ling, Jingwei and Javid, Usman A. and Xue, Shixin and He, Yang and Lopez-Rios, Raymond and Morin, Theodore J. and Wang, Heming and Shen, Boqiang and Zeng, Siwei and Zhu, Lin and Vahala, Kerry J. and Bowers, John E. and Lin, Qiang},
  title = {Integrated {Pockels} laser},
  journal = {Nature Communications},
  volume = {13},
  number = {1},
  pages = {5344},
  year = {2022},
  doi = {10.1038/s41467-022-33101-6},
  url = {https://www.nature.com/articles/s41467-022-33101-6},
}

@article{xue_pockels_2025,
  author = {Xue, Shixin and Li, Mingxiao and Lopez-Rios, Raymond and Ling, Jingwei and Gao, Zhengdong and Hu, Qili and Qiu, Tian and Staffa, Jeremy and Chang, Lin and Wang, Heming and Xiang, Chao and Bowers, John E. and Lin, Qiang},
  title = {{Pockels} laser directly driving ultrafast optical metrology},
  journal = {Light: Science \& Applications},
  volume = {14},
  number = {1},
  pages = {209},
  year = {2025},
  doi = {10.1038/s41377-025-01872-4},
  url = {https://www.nature.com/articles/s41377-025-01872-4},
}

@article{snigirev_ultrafast_2023,
  author = {Snigirev, Viacheslav and Riedhauser, Annina and Lihachev, Grigory and Churaev, Mikhail and Riemensberger, Johann and Wang, Rui Ning and Siddharth, Anat and Huang, Guanhao and Möhl, Charles and Popoff, Youri and Drechsler, Ute and Caimi, Daniele and Hönl, Simon and Liu, Junqiu and Seidler, Paul and Kippenberg, Tobias J.},
  title = {Ultrafast tunable lasers using lithium niobate integrated photonics},
  journal = {Nature},
  volume = {615},
  number = {7952},
  pages = {411--417},
  year = {2023},
  doi = {10.1038/s41586-023-05724-2},
  url = {https://www.nature.com/articles/s41586-023-05724-2},
}

@article{yuan_correlated_2022,
  author = {Yuan, Zhiquan and Wang, Heming and Liu, Peng and Li, Bohan and Shen, Boqiang and Gao, Maodong and Chang, Lin and Jin, Warren and Feshali, Avi and Paniccia, Mario and Bowers, John and Vahala, Kerry},
  title = {Correlated self-heterodyne method for ultra-low-noise laser linewidth measurements},
  journal = {Optics Express},
  volume = {30},
  number = {14},
  pages = {25147--25161},
  year = {2022},
  doi = {10.1364/OE.458109},
  url = {https://opg.optica.org/oe/abstract.cfm?uri=oe-30-14-25147},
}

@article{castro_integrated_2025,
  author = {Castro, Joshua E. and Nolasco-Martinez, Eber and Pintus, Paolo and Zhang, Zeyu and Shen, Boqiang and Morin, Theodore and Thiel, Lillian and Steiner, Trevor J. and Lewis, Nicholas and Patel, Sahil D. and Bowers, John E. and Weld, David M. and Moody, Galan},
  title = {Integrated mode-hop-free tunable lasers at 780 nm for chip-scale classical and quantum photonic applications},
  journal = {{APL} Photonics},
  volume = {10},
  number = {3},
  pages = {036102},
  year = {2025},
  doi = {10.1063/5.0232377},
  url = {https://doi.org/10.1063/5.0232377},
}

@inproceedings{luntadila_tunable_2023,
  author = {Luntadila Lufungula, Isaac and Mayor, Felix M. and
            Herrmann, Jason F. and Park, Taewon and
            Stokowski, Hubert S. and Hwang, Alexander Y. and
            Op de Beeck, Camiel and Atalar, Okan and
            Jiang, Wentao and Kuyken, Bart and Safavi-Naeini, Amir H.},
  title = {Tunable dual wavelength laser on thin film lithium niobate},
  booktitle = {2023 IEEE Photonics Conference (IPC)},
  pages = {MD3.5},
  year = {2023},
  publisher = {IEEE},
  doi = {10.1109/IPC57732.2023.10360651},
  url = {https://doi.org/10.1109/IPC57732.2023.10360651},
}

@article{xie_aluminum_2025,
author = {Xie, Hao and Wang, Yubo and Zhou, Yiyu and Holguín-Lerma, Jorge A. and Yang, Guangcanlan and Shi, Zhimin and Tang, Hong X.},
title = {Aluminum nitride photonics for high-speed electro-optical tuning of self-injection-locked laser},
journal = {Optics Letters},
volume = {50},
number = {18},
pages = {5913--5916},
year = {2025},
doi = {10.1364/OL.571314},
url = {https://opg.optica.org/ol/abstract.cfm?uri=ol-50-18-5913},
}

@article{corato-zanarella_widely_2023,
  author = {Corato-Zanarella, Mateus and Gil-Molina, Andres and Ji, Xingchen and Shin, Min Chul and Mohanty, Aseema and Lipson, Michal},
  title = {Widely tunable and narrow-linewidth chip-scale lasers from near-ultraviolet to near-infrared wavelengths},
  journal = {Nature Photonics},
  volume = {17},
  number = {2},
  pages = {157--164},
  year = {2023},
  doi = {10.1038/s41566-022-01120-w},
  url = {https://www.nature.com/articles/s41566-022-01120-w},
}

@article{winkler_widely_2024,
  author = {Winkler, Lisa V. and Gerritsma, Kirsten and van Rees, Albert and Schrinner, Philip P. J. and Hoekman, Marcel and Dekker, Ronald and do Nascimento, Jr., Adriano R. and van der Slot, Peter J. M. and Nölleke, Christian and Boller, Klaus-J.},
  title = {Widely tunable and narrow-linewidth hybrid-integrated diode laser at 637 nm},
  journal = {Optics Express},
  volume = {32},
  number = {17},
  pages = {29710--29720},
  year = {2024},
  doi = {10.1364/OE.523985},
  url = {https://opg.optica.org/oe/abstract.cfm?uri=oe-32-17-29710},
}

@article{clementi_chip-scale_2023,
  author = {Clementi, Marco and Nitiss, Edgars and Liu, Junqiu and Durán-Valdeiglesias, Elena and Belahsene, Sofiane and Debrégeas, Hélène and Kippenberg, Tobias J. and Brès, Camille-Sophie},
  title = {A chip-scale second-harmonic source via self-injection-locked all-optical poling},
  journal = {Light: Science \& Applications},
  volume = {12},
  number = {1},
  pages = {296},
  year = {2023},
  doi = {10.1038/s41377-023-01329-6},
  url = {https://www.nature.com/articles/s41377-023-01329-6},
}

@article{yu_resonator-enhanced_2026,
  author = {Yu, Di and Geng, Zhaoting and Huang, Yuhao and Tong, Yitian and Xia, Yu and Liu, Mingfei and Huang, Yaoran and Xiang, Chao},
  title = {Resonator-enhanced distributed {Bragg} reflector lasers},
  journal = {Light: Science \& Applications},
  volume = {15},
  number = {1},
  pages = {142},
  year = {2026},
  doi = {10.1038/s41377-026-02249-x},
  url = {https://www.nature.com/articles/s41377-026-02249-x},
}

@article{corwin_frequency-stabilized_nodate,
  author = {Corwin, Kristan L. and Lu, Zheng-Tian and Hand, Carter F. and Epstein, Ryan J. and Wieman, Carl E.},
  title = {Frequency-stabilized diode laser with the {Zeeman} shift in an atomic vapor},
  journal = {Applied Optics},
  volume = {37},
  number = {15},
  pages = {3295--3298},
  year = {1998},
  doi = {10.1364/AO.37.003295},
  url = {https://opg.optica.org/ao/abstract.cfm?uri=ao-37-15-3295},
}

@article{sane_11_2012,
  author = {Sané, S. S. and Bennetts, S. and Debs, J. E. and Kuhn, C. C. N. and McDonald, G. D. and Altin, P. A. and Close, J. D. and Robins, N. P.},
  title = {11 {W} narrow linewidth laser source at 780 nm for laser cooling and manipulation of Rubidium},
  journal = {Optics Express},
  volume = {20},
  number = {8},
  pages = {8915--8919},
  year = {2012},
  doi = {10.1364/OE.20.008915},
  url = {https://opg.optica.org/oe/abstract.cfm?uri=oe-20-8-8915},
}

@article{guan_780nm_2025,
  author = {Guan, Xiaolei and Zhang, Jia and Gao, Xun and Wang, Yu and Shi, Tiantian and Chen, Jingbiao},
  title = {A 780 nm optical frequency standard based on diffuse laser cooled {$^{87}$Rb} atoms},
  journal = {Applied Physics Letters},
  volume = {126},
  number = {3},
  pages = {031104},
  year = {2025},
  doi = {10.1063/5.0250471},
  url = {https://doi.org/10.1063/5.0250471},
}

@article{isichenko_multi-laser_2026,
  author = {Isichenko, Andrei and Hunter, Andrew S. and Chauhan, Nitesh and Dickson, John R. and Nunley, T. Nathan and Bingaman, Josiah R. and Heim, David A. S. and Harrington, Mark W. and Liu, Kaikai and Kunz, Paul D. and Blumenthal, Daniel J.},
  title = {Multi-laser stabilization with an atomic-disciplined photonic integrated resonator},
  journal = {Optica Quantum},
  volume = {4},
  number = {3},
  pages = {232--240},
  year = {2026},
  doi = {10.1364/OPTICAQ.577668},
  url = {https://opg.optica.org/opticaq/abstract.cfm?uri=opticaq-4-3-232},
}

@article{isichenko_sub-hz_2024,
  author = {Isichenko, Andrei and Hunter, Andrew S. and Bose, Debapam and Chauhan, Nitesh and Song, Meiting and Liu, Kaikai and Harrington, Mark W. and Blumenthal, Daniel J.},
  title = {Sub-{Hz} fundamental, sub-{kHz} integral linewidth self-injection locked 780 nm hybrid integrated laser},
  journal = {Scientific Reports},
  volume = {14},
  number = {1},
  pages = {27015},
  year = {2024},
  publisher = {Nature Publishing Group},
  doi = {10.1038/s41598-024-76699-x},
  url = {https://www.nature.com/articles/s41598-024-76699-x},
  copyright = {2024 The Author(s)},
  issn = {2045-2322},
  urldate = {2026-05-08},
}

@article{kitching_chip-scale_2018,
  author = {Kitching, John},
  title = {Chip-scale atomic devices},
  journal = {Applied Physics Reviews},
  volume = {5},
  number = {3},
  pages = {031302},
  year = {2018},
  month = aug,
  doi = {10.1063/1.5026238},
  url = {https://doi.org/10.1063/1.5026238},
  issn = {1931-9401},
  urldate = {2026-05-21},
}

@article{siddharth_ultrafast_2025,
  author = {Siddharth, Anat and Bianconi, Simone and Wang, Rui Ning and Qiu, Zheru and Voloshin, Andrey S. and Bereyhi, Mohammad J. and Riemensberger, Johann and Kippenberg, Tobias J.},
  title = {Ultrafast tunable photonic-integrated extended-{DBR} {Pockels} laser},
  journal = {Nature Photonics},
  volume = {19},
  number = {7},
  pages = {709--717},
  year = {2025},
  doi = {10.1038/s41566-025-01687-0},
  url = {https://www.nature.com/articles/s41566-025-01687-0},
  copyright = {2025 The Author(s), under exclusive licence to Springer Nature Limited},
  issn = {1749-4893},
  urldate = {2025-06-09},
}

@article{zhu_integrated_2021,
  author = {Zhu, Di and Shao, Linbo and Yu, Mengjie and Cheng, Rebecca and Desiatov, Boris and Xin, C. J. and Hu, Yaowen and Holzgrafe, Jeffrey and Ghosh, Soumya and Shams-Ansari, Amirhassan and Puma, Eric and Sinclair, Neil and Reimer, Christian and Zhang, Mian and Lončar, Marko},
  title = {Integrated photonics on thin-film lithium niobate},
  journal = {Advances in Optics and Photonics},
  volume = {13},
  number = {2},
  pages = {242--352},
  year = {2021},
  month = jun,
  publisher = {Optica Publishing Group},
  doi = {10.1364/AOP.411024},
  url = {https://opg.optica.org/aop/abstract.cfm?uri=aop-13-2-242},
  copyright = {© 2021 Optical Society of America},
  issn = {1943-8206},
  urldate = {2026-05-17},
}

@article{martin_compact_2018,
  author = {Martin, Kyle W. and Phelps, Gretchen and Lemke, Nathan D. and Bigelow, Matthew S. and Stuhl, Benjamin and Wojcik, Michael and Holt, Michael and Coddington, Ian and Bishop, Michael W. and Burke, John H.},
  title = {Compact {Optical} {Atomic} {Clock} {Based} on a {Two}-{Photon} {Transition} in {Rubidium}},
  journal = {Physical Review Applied},
  volume = {9},
  number = {1},
  pages = {014019},
  year = {2018},
  month = jan,
  publisher = {American Physical Society},
  doi = {10.1103/PhysRevApplied.9.014019},
  url = {https://link.aps.org/doi/10.1103/PhysRevApplied.9.014019},
  urldate = {2026-06-07},
}

@article{bluvstein_logical_2024,
  author = {Bluvstein, Dolev and Evered, Simon J. and Geim, Alexandra A. and Li, Sophie H. and Zhou, Hengyun and Manovitz, Tom and Ebadi, Sepehr and Cain, Madelyn and Kalinowski, Marcin and Hangleiter, Dominik and Bonilla Ataides, J. Pablo and Maskara, Nishad and Cong, Iris and Gao, Xun and Sales Rodriguez, Pedro and Karolyshyn, Thomas and Semeghini, Giulia and Gullans, Michael J. and Greiner, Markus and Vuletić, Vladan and Lukin, Mikhail D.},
  title = {Logical quantum processor based on reconfigurable atom arrays},
  journal = {Nature},
  volume = {626},
  number = {7997},
  pages = {58--65},
  year = {2024},
  month = feb,
  publisher = {Nature Publishing Group},
  doi = {10.1038/s41586-023-06927-3},
  url = {https://www.nature.com/articles/s41586-023-06927-3},
  copyright = {2023 The Author(s)},
  issn = {1476-4687},
  urldate = {2026-06-07},
}

@article{lu_periodically_2019,
  author = {Lu, Juanjuan and Surya, Joshua B. and Liu, Xianwen and Bruch, Alexander W. and Gong, Zheng and Xu, Yuntao and Tang, Hong X.},
  title = {Periodically poled thin-film lithium niobate microring resonators with a second-harmonic generation efficiency of 250,000\%/{W}},
  journal = {Optica},
  volume = {6},
  number = {12},
  pages = {1455--1460},
  year = {2019},
  month = dec,
  publisher = {Optica Publishing Group},
  doi = {10.1364/OPTICA.6.001455},
  url = {https://opg.optica.org/optica/abstract.cfm?uri=optica-6-12-1455},
  copyright = {© 2019 Optical Society of America},
  issn = {2334-2536},
  urldate = {2026-06-07},
}

@article{smith_theory_1970,
  author = {Smith, Richard G.},
  title = {Theory of intracavity optical second-harmonic generation},
  journal = {IEEE Journal of Quantum Electronics},
  volume = {6},
  number = {4},
  pages = {215--223},
  year = {1970},
  month = apr,
  doi = {10.1109/JQE.1970.1076440},
  url = {https://ieeexplore.ieee.org/document/1076440/},
  issn = {1558-1713},
  urldate = {2025-10-27},
}

@incollection{nurmikko_intracavity_2003,
  author = {Risk, W. P. and Gosnell, T. R. and Nurmikko, A. V.},
  title = {Intracavity {SHG} and {SFG}},
  booktitle = {Compact {Blue}-{Green} {Lasers}},
  pages = {223--262},
  chapter = {5},
  year = {2003},
  publisher = {Cambridge University Press},
  address = {Cambridge},
  isbn = {978-0-521-62318-6},
  doi = {10.1017/CBO9780511606502.006},
  url = {https://www.cambridge.org/core/books/compact-bluegreen-lasers/intracavity-shg-and-sfg/FD78B8149F6ADF6E052135137BC32A85},
  urldate = {2026-04-13},
}

@article{luEmergingIntegratedLaser2024b,
  author = {Lu, Xiyuan and Chang, Lin and Tran, Minh A. and Komljenovic, Tin and Bowers, John E. and Srinivasan, Kartik},
  title = {Emerging Integrated Laser Technologies in the Visible and Short Near-Infrared Regimes},
  journal = {Nature Photonics},
  volume = {18},
  number = {10},
  pages = {1010--1023},
  year = {2024},
  month = oct,
  publisher = {Nature Publishing Group},
  doi = {10.1038/s41566-024-01529-5},
  issn = {1749-4893},
  urldate = {2026-06-30},
  copyright = {2024 This is a U.S. Government work and not under copyright protection in the US; foreign copyright protection may apply},
}

@article{wangUltrafastHighFidelityState2025,
  author = {Wang, Jian and Huang, Dong-Yu and Zhou, Xiao-Long and Shen, Ze-Min and He, Si-Jian and Huang, Qi-Yang and Liu, Yi-Jia and Li, Chuan-Feng and Guo, Guang-Can},
  title = {Ultrafast {{High-Fidelity State Readout}} of {{Single Neutral Atom}}},
  journal = {Physical Review Letters},
  volume = {134},
  number = {24},
  pages = {240802},
  year = {2025},
  month = jun,
  doi = {10.1103/PhysRevLett.134.240802},
  issn = {0031-9007, 1079-7114},
  urldate = {2026-07-09},
}

@article{thiel2026wafer,
  author = {Thiel, L. and Shen, B. and Venneberg, J. R. and Guidry, M. A.
            and Arnaud, N. and Slater, A. and Wang, L. and Li, X.
            and Castro, J. and Pang, Y. and Meunier, M. and Patel, S. D.
            and Shen, Y. and Morin, T. and Kudelin, I. and Song, B.
            and Asawa, K. and Bowers, J. E. and Vahala, K.
            and Mavalvala, N. and Yin, X. and Bowers, S.
            and Tran, M. A. and Komljenovic, T. and Moody, G.},
  title = {A Wafer-Scale Heterogeneous {III-V}-on-Silicon Nitride Quantum Photonic Platform},
  journal = {arXiv preprint arXiv:2605.17738},
  year = {2026}
}

@article{zhangPhotonicIntegrationPlatform2023a,
  author = {Zhang, Zeyu and Shen, Boqiang and Tran, Minh A. and Lee, Woonghee and Asawa, Kaustubh and Kim, Glenn and Shen, Yang and Morin, Theodore J. and Malik, Aditya and Bowers, John E. and Komljenovic, Tin and Zhang, Chong},
  title = {Photonic Integration Platform for Rubidium Sensors and Beyond},
  journal = {Optica},
  volume = {10},
  number = {6},
  pages = {752--753},
  year = {2023},
  month = jun,
  publisher = {Optica Publishing Group},
  doi = {10.1364/OPTICA.494716},
  issn = {2334-2536},
  urldate = {2026-05-20},
  copyright = {\copyright{} 2023 Optica Publishing Group},
}

@article{flanniganStudyFundamentalWave2021,
  author = {Flannigan, Liam and Kashak, Tyler and Xu, Chang-Qing},
  title = {Study of Fundamental Wave Depletion in Intracavity Second Harmonic Generation},
  journal = {Optics Express},
  volume = {29},
  number = {5},
  pages = {6810--6823},
  year = {2021},
  month = mar,
  publisher = {Optica Publishing Group},
  doi = {10.1364/OE.417191},
  issn = {1094-4087},
  urldate = {2025-10-27},
  copyright = {\copyright{} 2021 Optical Society of America},
}

@article{moustakasOptoelectronicDevicePhysics2017,
  author = {Moustakas, Theodore D and Paiella, Roberto},
  title = {Optoelectronic Device Physics and Technology of Nitride Semiconductors from the {{UV}} to the Terahertz},
  journal = {Reports on Progress in Physics},
  volume = {80},
  number = {10},
  pages = {106501},
  year = {2017},
  month = sep,
  publisher = {IOP Publishing},
  doi = {10.1088/1361-6633/aa7bb2},
  issn = {0034-4885},
  urldate = {2026-07-09},
}

@article{holzgrafe_relaxation_2024,
  author = {Holzgrafe, Jeffrey and Puma, Eric and Cheng, Rebecca and Warner, Hana and Shams-Ansari, Amirhassan and Shankar, Raji and Lon{\v{c}}ar, Marko},
  title = {Relaxation of the electro-optic response in thin-film lithium niobate modulators},
  journal = {Optics Express},
  volume = {32},
  number = {3},
  pages = {3619--3631},
  year = {2024},
  month = jan,
  doi = {10.1364/OE.507536},
  url = {https://doi.org/10.1364/OE.507536},
}

@article{ferrari_precision_2003,
  author  = {Ferrari, G. and Cancio, P. and Drullinger, R.
             and Giusfredi, G. and Poli, N. and Prevedelli, M.
             and Toninelli, C. and Tino, G. M.},
  title   = {Precision Frequency Measurement of Visible
             Intercombination Lines of Strontium},
  journal = {Physical Review Letters},
  year    = {2003},
  volume  = {91},
  number  = {24},
  pages   = {243002},
  doi     = {10.1103/PhysRevLett.91.243002}
}

@article{siddharth2025narrow,
  author = {Siddharth, A. and Gardner, A. B. and Ji, X.
            and Hulyal, S. U. and Reichler, M. S.
            and Attanasio, A. and Riemensberger, J.
            and Bhave, S. A. and Volet, N.
            and Bianconi, S. and Kippenberg, T. J.},
  title = {Narrow-Linewidth, Piezoelectrically Tunable Photonic Integrated Blue Laser},
  journal = {arXiv preprint arXiv:2508.02568},
  year = {2025}
}

@article{leeLaserFrequencyStabilization2023,
  title = {Laser Frequency Stabilization in the {$10^{-14}$} Range via Optimized Modulation Transfer Spectroscopy on the {$^{87}$Rb} {$D_2$} Line},
  author = {Lee, Sanglok and Moon, Geol and Park, Sang Eon and Hong, Hyun-Gue and Lee, Jae Hoon and Seo, Sangwon and Kwon, Taeg Yong and Lee, Sang-Bum},
  year = {2023},
  month = feb,
  journal = {Optics Letters},
  volume = {48},
  number = {4},
  pages = {1020--1023},
  doi = {10.1364/OL.480178}
}

@article{corato-zanarella_lowswap_2026,
  title = {Low-{SWaP}, Tunable, and Narrow-Linewidth Laser Systems for Deployable Quantum Technologies},
  author = {Corato-Zanarella, Mateus and Lommel, Matthias and Mayzlin, Dimitri and Lopes, Emilio and Nojic, Jovana and Eras, Alexander and N{\"o}lleke, Christian and Globisch, Bj{\"o}rn and Haimberger, Christopher},
  year = {2026},
  month = mar,
  journal = {Optics Express},
  volume = {34},
  number = {5},
  pages = {9308--9317},
  doi = {10.1364/OE.588093}
}

@article{shi2026integrated,
  title = {Integrated polarization-entangled photon source for wavelength-multiplexed quantum networks},
  author = {Shi, Xiaodong and Li, Yue and Du, Jinyi and Zhou, Lin and Yang, Ran and
            Lim, En Teng and Mohanraj, Sakthi Sanjeev and Zhao, Mengyao and
            Chen, Xu and Wang, Xiaojie and Wu, Guangxing and Hao, Hao and
            Dhyani, Veerendra and Wang, Sihao and Ling, Alexander and Zhu, Di},
  journal = {Nature Communications},
  volume = {17},
  number = {1},
  pages = {9778},
  year = {2026},
  month = aug,
  doi = {10.1038/s41467-026-76631-z}
}

@article{larocque_sub-khz_2026,
  author = {Larocque, Hugo and Yuan, Zhuoya and Li, Zihan and Scarioni, Giovanni and Schoels, Annika and Sun, Jiale and Siddharth, Anat and Ou, Xin and Bianconi, Simone and Kippenberg, Tobias J.},
  title = {{Sub-kHz} linewidth integrated {extended-DBR} {Pockels} lasers using lithium tantalate},
  journal = {arXiv preprint arXiv:2609.14777},
  year = {2026},
  url = {https://arxiv.org/abs/2609.14777}
}

@article{evered_high-fidelity_2023,
  author = {Evered, Simon J. and Bluvstein, Dolev and Kalinowski, Marcin and Ebadi, Sepehr and Manovitz, Tom and Zhou, Hengyun and Li, Sophie H. and Geim, Alexandra A. and Wang, Tout T. and Maskara, Nishad and Levine, Harry and Semeghini, Giulia and Greiner, Markus and Vuleti{\'c}, Vladan and Lukin, Mikhail D.},
  title = {High-fidelity parallel entangling gates on a neutral-atom quantum computer},
  journal = {Nature},
  volume = {622},
  number = {7982},
  pages = {268--272},
  year = {2023},
  doi = {10.1038/s41586-023-06481-y},
  url = {https://www.nature.com/articles/s41586-023-06481-y},
}

@article{stray_quantum_2022,
  author = {Stray, Ben and Lamb, Andrew and Kaushik, Aisha and Vovrosh, Jamie and Rodgers, Anthony and Winch, Jonathan and Hayati, Farzad and Boddice, Daniel and Stabrawa, Artur and Niggebaum, Alexander and Langlois, Mehdi and Lien, Yu-Hung and Lellouch, Samuel and Roshanmanesh, Sanaz and Ridley, Kevin and de Villiers, Geoffrey and Brown, Gareth and Cross, Trevor and Tuckwell, George and Faramarzi, Asaad and Metje, Nicole and Bongs, Kai and Holynski, Michael},
  title = {Quantum sensing for gravity cartography},
  journal = {Nature},
  volume = {602},
  number = {7898},
  pages = {590--594},
  year = {2022},
  doi = {10.1038/s41586-021-04315-3},
  url = {https://www.nature.com/articles/s41586-021-04315-3},
}

@article{morgenweg_ramsey-comb_2014,
  author = {Morgenweg, Jonas and Barmes, Itan and Eikema, Kjeld S. E.},
  title = {Ramsey-comb spectroscopy with intense ultrashort laser pulses},
  journal = {Nature Physics},
  volume = {10},
  number = {1},
  pages = {30--33},
  year = {2014},
  doi = {10.1038/nphys2807},
  url = {https://www.nature.com/articles/nphys2807},
}

@article{hilton_mobile_2025,
  author = {Hilton, A. P. and Offer, R. F. and Klantsataya, E. and
            Scholten, S. K. and Bourbeau H{\'e}bert, N. and
            Billington, C. J. and Locke, C. and Perrella, C. and
            Nelligan, M. and Allison, J. W. and White, B. and
            Ahern, E. and Martin, K. W. and Beard, R. and
            Elgin, J. D. and Sparkes, B. M. and Luiten, A. N.},
  title = {Demonstration of a mobile optical clock ensemble at sea},
  journal = {Nature Communications},
  volume = {16},
  number = {1},
  pages = {6063},
  year = {2025},
  month = jul,
  doi = {10.1038/s41467-025-61140-2},
  url = {https://www.nature.com/articles/s41467-025-61140-2},
}

\end{document}